\documentclass[twocolumn,aps,showpacs,prl,preprintnumbers,bibnotes7pt]{revtex4}
\usepackage{epsfig}
\usepackage{graphicx}
\usepackage[countmax]{subfloat}

\usepackage{xcolor}     
\usepackage{soul}       
\begin{document}

\title{Universality of the three-body Efimov parameter at narrow Feshbach resonances}
\author{Sanjukta Roy$^1$}
\author{Manuele Landini$^{1}$}
\author{Andreas Trenkwalder$^{1}$}
\author{Giulia Semeghini$^1$}
\author{Giacomo Spagnolli$^1$}
\author{Andrea Simoni$^3$}
\author{Marco Fattori$^{1,2}$}
\author{Massimo Inguscio$^{1,2}$}
\author{Giovanni Modugno$^{1,2}$}

\affiliation{$^1$LENS and Dipartimento di Fisica e Astronomia, Universit\'a di Firenze, and
Istituto Nazionale di Ottica, CNR, 50019 Sesto Fiorentino, Italy} \affiliation{$^2$INFN, Sezione di
Firenze, 50019 Sesto Fiorentino, Italy} \affiliation{$^3$Institut de Physique de Rennes, UMR 6251,
CNRS and Universit\'e de Rennes 1, 35042 Rennes Cedex, France}

\date{\today}

\begin{abstract}
We measure the critical scattering length for the appearance of the first three-body bound state,
or Efimov three-body parameter, at seven different Feshbach resonances in ultracold $^{39}$K atoms.
We study both intermediate and narrow resonances, where the three-body spectrum is expected to be
determined by the non-universal coupling of two scattering channels. We observe instead
approximately the same universal relation of the three-body parameter with the two-body van der
Waals radius already found for broader resonances, which can be modeled with a single channel. This
unexpected observation suggests the presence of a new regime for three-body scattering at narrow
resonances.
\end{abstract}


\maketitle

Recent experiments on ultracold atoms with Feshbach resonances \cite{Kraemer,Potassium,Minardi,
Hulet, Hulet1, Khaykovich1, Khaykovich2,Jochim1, Hara1, Hara2,Jochim2, Ueda, Rb} have opened up a
new path to study the Efimov spectrum of the three-body bound states that arise in the presence of
resonant two-body interactions \cite{Efimov}. Under these conditions, the details of the underlying
forces become irrelevant, leading to a universal behavior of completely different systems. This
phenomenon was first described in the context of nuclear physics, but is now explored also in
atomic, molecular and condensed-matter systems \cite{Nielsen,Braaten,review,magnons}. The resonant
interaction is expected to give rise to a three-body potential scaling as $1/R^2$, where $R$ is the
hyperradius that parametrizes the moment of inertia of the system. This leads to an infinite series
of trimer states with a universal geometrical scaling for the binding energies. For a finite,
negative two-body scattering length $a$, the three-body potential has a long-range cutoff at
$R\simeq|a|$, and only a finite number of bound states exist. The critical scattering length $a_-$
for the appearance of the first Efimov state at the three-body threshold, often called the
three-body parameter, was expected to be the only parameter to be influenced by non-universal
physics, i.e. by the microscopic details of two or even three-body forces \cite{Efimov,Braaten}.
While a clear evidence of the universal scaling of the Efimov spectrum is still missing, recent
experiments on identical bosons suggested that also $a_-$ might be universal \cite{Berninger}. This
surprising result has been interpreted in a recent series of theoretical studies
\cite{Chin,Greene,Ueda1}. The underlying idea is that the sharp drop in the two-body interaction
potential at a distance of the order of the van der Waals radius $R_{vdW}$ results in an effective
barrier in the three-body potential at a comparable distance \cite{Ueda1}. This prevents the three
particles to come sufficiently close to explore non-universal features of the interactions at short
distances, and leads to a three-body parameter set by $R_{vdW}$ alone, $a_- \simeq -9.5\,R_{vdW}$
\cite{Berninger,Greene,Ueda1}.

However, this scenario is realized only for the broad Feshbach resonances studied so far in most
experiments, which can be described in terms of a single scattering channel, the so-called
open channel. For narrow resonances one must instead take into account the coupling of the open and
a second closed channel \cite{Feshbach}. It has been shown that in this case a new length scale
that depends on the details of the specific Feshbach resonance, the so-called intrinsic length
$R^*>R_{vdW}$, must be introduced to parameterize the two-body scattering. The three-body
potentials are also modified, with an expected deviation from the Efimovian dependence into
$1/(R^*R)$ for distances $R<R^*$ \cite{Petrov}. This tends to reduce the depth of the three-body
potential, and leads to the non-universal result $a_-=-12.90 R^*$ \cite{Petrov,Gogolin}, which is
much larger than that obtained for broad resonances. This prediction is valid only close to
resonance, where $|a|\gg R^*$. It is still unclear how $a_-$ scales in the intermediate regime of
$|a|\simeq R^*$ or generally for resonances of intermediate widths. Various general models have
been proposed \cite{Massignan,Pricoupenko1,Esry,Pricoupenko2,Zwerger,Zinner}, but they are either
not fully predictive, or give contradicting results.

In this Letter we address this problem by performing an experimental study of three-body collisions
in ultracold bosonic $^{39}$K atoms, where we determine the three-body parameter $a_-$ at
several Feshbach resonances of intermediate or narrow width. In particular, our measurements probe
for the first time the regime of very small resonance strengths, $s_{res}=0.956
R_{vdW}/R^*\simeq0.1$, where $R^*$ might be expected to be the relevant length-scale that
determines $a_-$. Surprisingly, we find values of $a_-$ that are around the same  $-9.5 R_{vdW}$
measured for broad resonances, suggesting the existence of a novel intermediate regime of
three-body scattering.

The investigation of closed-channel dominated Feshbach resonances is particularly favoured in
$^{39}$K, which has several resonances with moderate magnetic width $\Delta$ and relatively small
background scattering length $-a_{bg}\simeq$20-30~$a_0$ \cite{Chiara}. These parameters, together
with the difference of the magnetic moments of the closed and open channels, $\delta\mu$, determine
the intrinsic length $R^*=\hbar^2/(m a_{bg} \Delta \delta\mu)$ \cite{Feshbach}. In particular, we
investigated seven different resonances with $s_{res}$ in the range 0.1-2.8 in the three magnetic
sub-levels of the hyperfine ground state $F$=1 \cite{Chiara}. One of the broadest of those
resonances was already studied before \cite{Potassium}, and our new data clarifies an apparent
deviation from the universal behaviour.

A detailed description of the experimental set-up and methods for preparing Bose-Einstein
condensates of $ ^{39}$K atoms by direct evaporation is given elsewhere \cite{BEC39}. The
three-body parameter was determined by finding the maximum of the three-body loss coefficient $K_3$
in the region of negative $a$ at each Feshbach resonance, as in previous experiments
\cite{Kraemer,Potassium,Minardi, Hulet, Hulet1, Khaykovich1, Khaykovich2,Jochim1, Hara1,
Hara2,Jochim2, Ueda, Rb}. In the presence of three-body losses, both the atom number $N$ and
temperature $T$ evolve according to $dN/dt=-K_3\langle n^2\rangle N$ and $dT/dt=(K_3/3)\langle
n^2\rangle T$, where $\langle n^2\rangle=(1/N)\int n(\vec{x})^3d^3x$ is the mean square density
\cite{Weber}. The temperature increase is due to the preferential removal of atoms in the
high-density region around the trap center. The typical starting condition was a non-condensed
sample with 3-80$\times 10^{4}$ atoms in a temperature range of 20-400~nK, depending on the spin
channel and Feshbach resonance \cite{suppl}. The atoms were held in a purely optical trap (or in an
optical trap with an additional magnetic confinement, depending on the specific resonance) at
sufficiently low density to have a negligible mean-field interaction energy. Care was taken to have
a trap depth sufficiently large to avoid an evaporation associated to the heating. The samples were
initially prepared at small negative $a$ in proximity of the Feshbach resonances; the measurements
started 10~ms after the scattering length was ramped to the final value in about 2~ms.

\begin{figure}[ht]
\includegraphics[width=0.95\columnwidth] {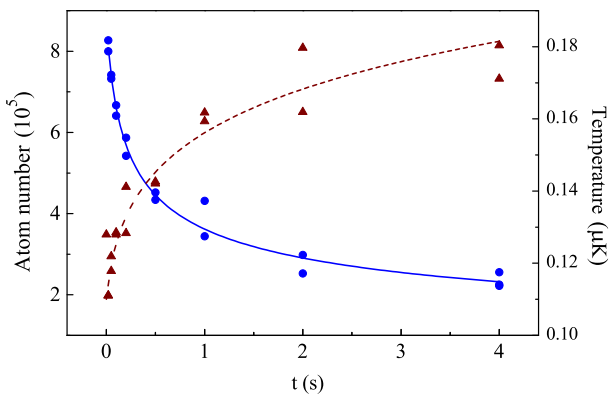}
\caption{Example of the time-evolution of the atom number (circles)
and temperature (triangles), fitted to Eq.1 (solid line) and Eq.2 (dashed line) to
determine the three-body loss-coefficient $K_3$.}
\label{fitting}
\end{figure}

Fig. \ref{fitting} shows a typical evolution of $N$ and $T$, as measured by absorption imaging
after a free expansion. They were simultaneously fitted with
\begin{eqnarray}N(t)= N_0/(1+\frac{3\beta^2}{\sqrt{27}}\frac{N_0^2}{T_0^3}K_3t)^{1/3}\,,\\
T(t)=T_0(1+ \frac{3\beta^2}{\sqrt{27}}\frac{N_0^2}{T_0^3}K_3t)^{1/9}\,.
\end{eqnarray}
Here $ N_{0} $ and $ T_{0} $ are the initial atom number and temperature, respectively, and
$\beta=(m\bar{\omega}^2/2\pi k_B)^{3/2}$, with $\bar{\omega}$ the mean trap frequency. In such a fit,
one-body losses were neglected, since they occur on a much longer timescale.

Crucial ingredients for a reliable measurement of the $K_3$ dependence on the scattering length
were an accurate calibration of the magnetic field $B$ and the use of a high-quality
coupled-channel (CC) model for $a(B)$, based on a large number of experimental observations for the
positions and widths of the Feshbach resonances \cite{Chiara,suppl}. The centers and widths of the
Feshbach resonances were redetermined in the present work, finding a good agreement with the
theoretical ones. An additional confirmation of the CC model was derived from a direct measurement
of the dimer binding energy at the two narrowest resonances by radio-frequency spectroscopy. The
magnetic field had a stability of better than 0.1~G, and was calibrated by means of radio-frequency
or microwave spectroscopy with an accuracy of 0.1~G. The inhomogeneity of $B$ across the atomic
samples was estimated to be less than 0.01 G in all cases.

\begin{figure}[ht!]
\includegraphics[width=0.95\columnwidth] {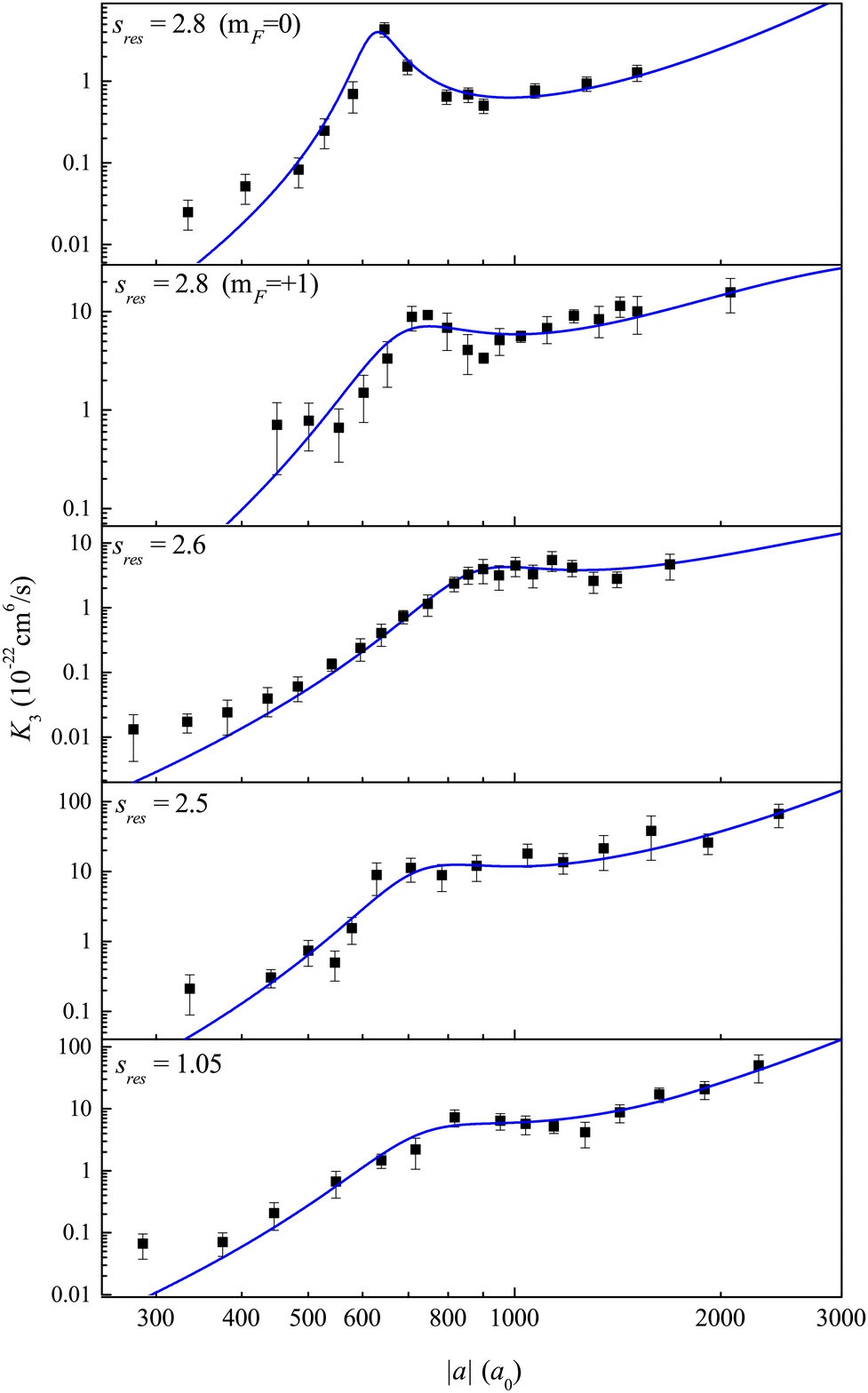}
\caption{Three-body loss rate measured in the proximity of five Feshbach resonances of intermediate strength
(see Table 1 for the assignment of the spin state). The experimental data (squares)
 is fitted to Eq.\ref{Universal_theory} incorporating the effect of unitarity at finite temperature (solid line).}
 \label{resonances1}
\end{figure}

\begin{figure}[ht!]
\includegraphics[width=0.95\columnwidth] {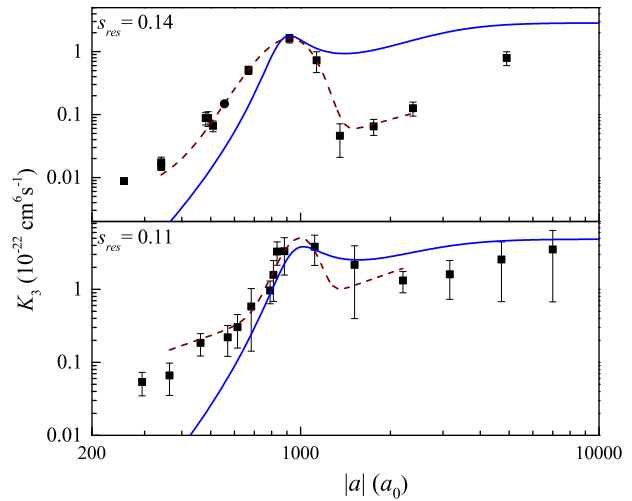}
\caption{Three-body loss rate measured in the proximity of two narrow Feshbach
resonances in the $m_F=0$ state. The experimental data (squares)
 is compared with Eq.\ref{Universal_theory}, using $\eta_-$=0.1 (solid line),
 and fitted with a Gaussian (dashed line) to determine $ a_{-} $ from the position of the
 loss maximum. }\label{resonances2}
\end{figure}

We observed for all Feshbach resonances a clear peak in $K_3$ in the region of
$|a|$=600-1000~$a_0$, as shown in Figs.\ref{resonances1}-\ref{resonances2}. We compared the
observations to the known relation for identical bosons at zero collision energy and in the
zero-range approximation, for $a<0$:
\begin{equation}
K_3(a)=4590 {3\hbar a^4
\over m}{{\text{sinh}(2\eta_-)}\over {\text{sin}^2[s_0\text{ln}(a/a_-)] +
\text{sinh}^2\eta_-}}\,. \label{Universal_theory}
\end{equation}
Here $s_0\simeq 1.00624$ is an universal constant and $\eta_-$ is the decay parameter which sets
the width of the Efimov resonance and incorporates short-range inelastic transitions to deeply
bound molecular states \cite{Braaten}. At the finite temperature of the experiment, there is a
limitation in the maximum observable $K_3$ set by unitarity at
$K_3^{max}=36\sqrt{3}\pi^2\hbar^5/(k_{B}T)^2m^3$ \cite{unitary,Petrov1}. Therefore, we used an
effective rate of the form $(1/K_3(a)+1/K_3^{max})^{-1}$. As shown in Fig.\ref{resonances1}, the
experimental $K_3(a)$ for the five broadest resonances is in good agreement with
Eq.\ref{Universal_theory}, besides a multiplicative factor of order 3 that can be justified with
the experimental uncertainty in the determination of the density \cite{suppl}. We extracted the
relevant parameters $a_-$ and $\eta_-$ with a fit.

Also the two narrowest Feshbach resonances feature a maximum in $K_3$ around -1000~$a_0$, as shown
in Fig.\ref{resonances2}. There is however a slower background variation of $K_3$ with $a$. It was
shown that for narrow resonances one should expect a slower evolution in the regime $|a|<R^*$, with
$K_3\propto |a|^{7/2}$ \cite{Petrov}, but also this behavior does not seem to reproduce the data.
Since these two resonances are in an excited spin state, there is in principle also a contribution
of two-body processes in the losses, which are expected to have a slower dependence on $a$
\cite{Feshbach}. While it was not possible to distinguish in a reliable way two- from three-body
losses in the experiment, we have verified that only the observed $K_3(a)$ far from the loss maxima
might be partially attributed to two-body losses \cite{suppl}. The low-$|a|$ tail at the Feshbach
resonance centered at $B_0=58.92$~G might also be affected by a nearby narrow $d$-wave resonance
\cite{suppl}. While the deviations of the measured $K_3$ from theory for these narrow resonances
will deserve further investigation, in the present work we identified the position of the Efimov
resonance with the peak of a Gaussian fit of the measured maximum in $K_3$ as shown in Fig. \ref{resonances2}.

A summary of our analysis is reported in Table \ref{tab:results}. For the calculation of $a(B)$, we
used the experimentally determined Feshbach resonance centers $B_0^{exp}$ and the resonance widths
and the background scattering lengths from the CC model. The uncertainties in $B_0^{exp}$ include
those in the calibration of $B$ and in the determination of $B_0$ from the loss resonances.
Particular care was put in the determination of $B_0$ for the two narrowest resonances, where we
found a rather good agreement between independent measurements of the atom losses and of the
binding energy \cite{suppl}. The uncertainties in $a_-$ include the statistical uncertainties from
the fit of the $K_{3} $ data and from the determination of $a(B)$. For the two narrowest
resonances, the dominant source of uncertainty comes from the determination of $B_0$. The reported
values of $R^*$ are determined from the on-resonance predictions of our CC model
\cite{Chiara,suppl}. We observe a whole range of values of $\eta_-$ for the different Efimov
resonances; this is probably a consequence of the different measurement temperatures, but possibly
also of the non-universal nature of $\eta_-$.

\begin{table}[t]

\centering \caption{Theoretical and experimental parameters for the three-body resonances at
Feshbach resonances in the $m_F$ spin channels: measured resonance center $B_0^{exp}$; intrinsic
length $R^*$ and strength $s_{res}$ of the Feshbach resonances from the CC model; measured
three-body parameter $a_-$ and decay parameter $\eta_-$; initial temperature $T$ of the atomic
sample. For $^{39}$K, $R_{vdW}$=64.49~$a_0$.}

\begin{tabular*}{8.5cm}{c c c c c c c c}
\hline \hline \rule[-2ex]{0pt}{3.5ex} m$ _{F}$ & B$_0^{exp}$(G) & $R^*(a_0)$  & $s_{res}$ & $-a_- (a_0)$ & $\eta_-$  & $T$ (nK) \\
\hline \rule[-1ex]{0pt}{3.5ex} 0   & 471.0 (4) & 22  & 2.8 & 640 (100)  & 0.065 (11)& 50 (5)\\
\rule[-1ex]{0pt}{3.5ex} +1  & 402.6 (2) & 22  & 2.8 & 690 (40) & 0.145 (12)& 90 (6)\\
\rule[-1ex]{0pt}{3.5ex} -1  & 33.64 (15) & 23 & 2.6 & 830 (140)  & 0.204 (10) & 120 (10)\\
\rule[-1ex]{0pt}{3.5ex} -1 & 560.72 (20) & 24  & 2.5 & 640 (90) & 0.22 (2)& 20 (7)  \\
\rule[-1ex]{0pt}{3.5ex} -1 & 162.35 (18) & 59  & 1.1 & 730 (120 & 0.26 (5)  & 40 (5)\\
\rule[-1ex]{0pt}{3.5ex} 0  & 65.67 (5) & 456  & 0.14 & 950 (250) & & 330 (30)\\
\rule[-1ex]{0pt}{3.5ex} 0  & 58.92 (3) & 556 & 0.11 & 950(150) & &  400 (80)\\

\hline \hline
\end{tabular*}
\label{tab:results}
\end{table}
\vspace{5mm}

A comparison of the results in Table 1 leads to the striking conclusion that the three-body
parameter $a_-$ stays around values of the order of -10~$R_{vdW}$ for all the Feshbach resonances
explored in $^{39}$K, including the ones with $R^*$ as large as $\sim 600~a_0$, hence much larger
than $R_{vdW}$. We note that in the earlier measurement at the resonance in the $m_F$=1 state
\cite{Potassium} a similar value of $a_-$ was found, but we cannot confirm the additionally
observed feature at 1500~$a_0$. We suspect this was an artefact of the analysis of the limited
time-dependent data.

\begin{figure}[ht]
\includegraphics[width=0.9\columnwidth] {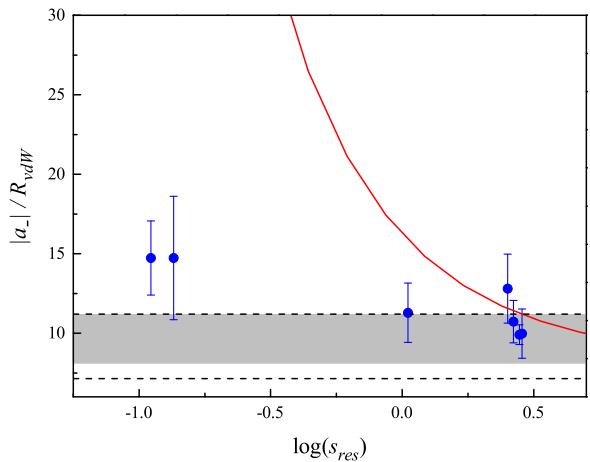}
\caption{(color online) Ratio of the measured three-body parameter to the van der Waals
radius for $^{39}$K atoms as a function of the strength of the Feshbach resonance
(blue circles) and comparison to the predictions of theoretical models in Ref.\cite{Greene}
(gray shaded region) and Ref.\cite{Zwerger} (solid line).
The dashed lines show the maximum and minimum of the scatter in the experimental data for the three-body
parameters measured at broad Feshbach resonances in other atomic species.}
\label{TBP}
\end{figure}

Fig.\ref{TBP} shows the measured $|a_-|/R_{vdW}$ as a function of $s_{res}$. One notes just a
moderate deviation of our data from the mean value 9.73(3) measured for open-channel dominated
resonances \cite{Greene,Berninger, Grimm, Rb}, and also for other intermediate resonances
\cite{Hulet, Hulet1, Khaykovich1, Khaykovich2,Berninger, Grimm}. This observation is far from the
already mentioned prediction for narrow resonances \cite{Petrov,Gogolin}, which indicates that the
Efimov resonances should appear at scattering lengths that are multiples of $a_-=-12.9R^*$ by a
factor exp$(\pi/s_0)\simeq22.7$. One might note that this result is expected to be valid only in
the limit of a scattering length larger than any other length scale, $|a|\gg R^* \gg |a_{bg}|$,
where the three-body potential at large hyperradii $R>R^*$ has an Efimovian character
\cite{Petrov}. The present experiment does not access this extreme limit but is in an intermediate
regime also for the two narrowest resonances, which show indeed $R^*\simeq |a_-|$.

Other models for the three-body physics at Feshbach resonances of intermediate strength have been
proposed \cite{Massignan,Pricoupenko1,Esry,Pricoupenko2,Zwerger,Zinner}. The specific problem of
connecting the results for the three-body parameter in the open-channel dominated regime, where
$a_-$ is determined by $R_{vdW}$, and the closed channel limit, where it is $R^*$ which sets the
scale for $a_-$, has been addressed recently \cite{Zwerger,Zinner}, finding however considerably
different results. In particular, the model of Ref.\cite{Zwerger} predicts that a crossover between
the two regimes of broad and narrow resonances would take place around $s_{res}\simeq1$, as shown
in Fig.\ref{TBP}. Additionally, the regime of $a_-=-12.9R^*$ should be reached only for excited
Efimov states, while the first one has a slightly smaller $a_-=-10.3R^*$. Although an increase of
$|a_-|$ with decreasing $s_{res}$ might be present in the experimental data, there is a clear
disagreement with such predictions. Experiments on $^7$Li and $^{133}$Cs have also measured similar
values for $a_-$ at three intermediate resonances with $s_{res}=0.5-1$ \cite{Khaykovich1,
Khaykovich2,Berninger,Hulet1}, indicating that this behavior might not be peculiar of $^{39}$K.

We note that for the two narrowest resonances $|a_-|$ is only a factor of two larger than $R^*$.
This observation seems to indicate that the three-body potential can support a bound state that
resides only in the region with hyperradius $R\leq 2R^*$. This is a regime that was not accessible
in previous one-channel models, and a multi-channel approach will be presumably necessary to model
the experimental observations.

In conclusion, our study showed an apparent universal behaviour of the three-body parameter on
several different Feshbach resonances of the same atomic species, down to a resonance strength
$s_{res}\simeq 0.1$. This gives important information on the three-body physics in this
narrow-resonance regime, where one expects a combined role of the open and closed molecular
channels. Our results will provide a benchmark for three-body multichannel models. By employing a
narrower, low-field Feshbach resonance in $^{39}$K \cite{Chiara}, further experiments probing a
regime of even smaller $s_{res}\simeq0.01$ might be possible in the future.

We thank R. Schmidt, M. Zaccanti and W. Zwerger for helpful discussions. This work was supported by
the INFM (MICRA collaboration), the European Research Council (grants 203479 and 247371), the
Agence Nationale de la Recherche (Contract No. ANR-12-BS04-0020-01) and the Italian Ministry for
Research (PRIN 2009FBKLNN).

\section{Supplementary material}

\subsection{Measurement of the Feshbach resonances parameters}
\begin{figure*}[ht]
\includegraphics[width=1.9\columnwidth] {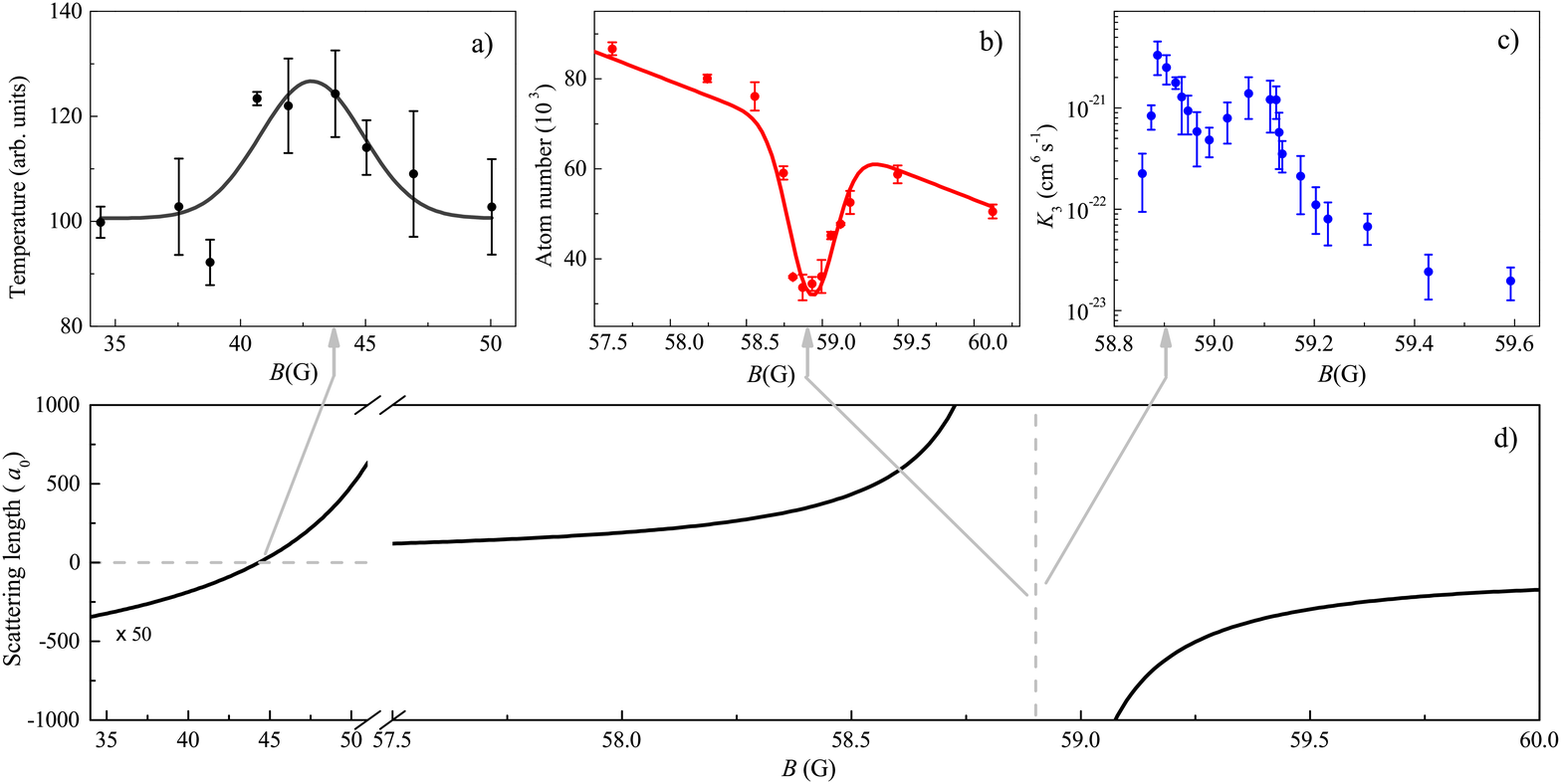}
\caption{Example of the calibration measurements for the narrow Feshbach resonance centered at
$B_0\simeq$58.9~G in the $m_F$=0 state. a) The magnetic-field position of the zero crossing was
determined by evaporation measurements to be at $B_{zc}$=43(2)~G. b) The position of the resonance
center was determined via loss measurements to be at $B_0$=58.9(2)~G. c) The center was
independently determined from the three-body loss data to be at $B_0$=58.92(3)(10)~G, where the
first uncertainty is statistical, and the second one is the accuracy in the absolute calibration of
$B$. Lines are Gaussian fits to the data. d) The scattering length vs the magnetic field from the
CC model of Ref.\cite{Chiara} predicts the zero crossing at $B_{zc}$=44.3(2)~G and the center at
$B_0$=58.8(1)~G. To make the zero-crossing more visible, the scattering length in that region was
multiplied by a factor 50.} \label{complete}
\end{figure*}

Given the importance of the quality of the coupled-channels (CC) model to predict a reliable
$a(B)$, we confirmed experimentally several of its predictions for the seven Feshbach resonances
investigated in this work. One example of the typical calibration measurements is shown in
Fig.\ref{complete} for the narrow resonance at $B_0\simeq$58.9~G in the $m_F$=0 state. For example,
the centers of the Feshbach resonances were experimentally determined, by measuring the atom losses
versus the magnetic field after a constant waiting time. The resonance centers were identified with
the loss maxima. Also the positions of the zero-crossings of $a$ were determined as minima in the
elastic collisional rate, after a forced evaporation procedure. The zero crossings were identified
with the maxima in the temperature. In all cases we found a very good agreement within the typical
experimental and theoretical uncertainties, as shown for example for the resonance centers in Table
\ref{tabsi:results}. The uncertainties in the experiment are of the order of 0.1~G for the centers and 1~G for the
widths. The corresponding uncertainties for the CC calculations are of the order of 0.1~G.

Note that the theoretical parameters for the Feshbach resonances are slightly different from those
previously reported \cite{Chiara}, which considered only $s$-wave scattering; the present
calculations include instead also $d$- and $g$-wave scattering. In Table \ref{tabsi:results} we also report a narrow
$d$-wave resonance, also predicted by the CC model, in the vicinity of the narrow resonance at
$B_0\simeq$58.9~G. Its presence limited the range of accessible scattering lengths to $|a|\simeq
200 a_0$.

We used particular care in determining the centers of the two narrow resonances, where the
interesting region of $|a|\simeq R^*$ appears at detunings $B-B_0$ of the order of 0.1~G, which is
also our typical accuracy in the calibration of the magnetic field. In these cases, the resonance
center was determined also from the measurement of the three-body loss coefficient on both sides of
the resonance, thus achieving a precision as low as 0.03~G. One measurement is shown in Fig.\ref{complete}c. In
absence of a detailed model of the loss rates very close to the resonance center, we performed a
Gaussian fit of the loss maximum, taking the center as $B_0$ and the halfwidth at 1/$e^2$ as
the uncertainty. The values for $B_0$ measured via the plain losses and $K_3$ are in good agreement
between themselves and with the theory, with a typical deviation below 0.1~G.

\begin{figure}[ht]
\includegraphics[width=0.9\columnwidth] {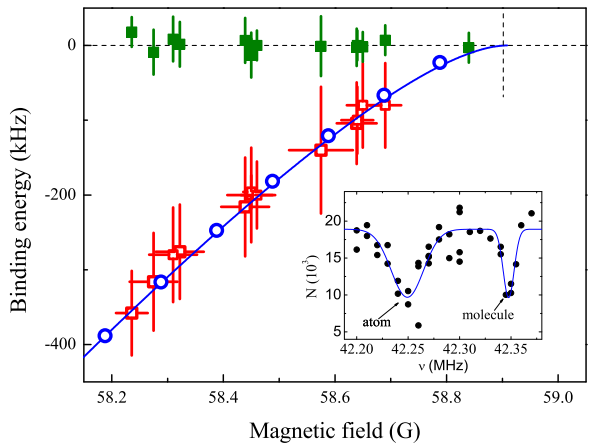}
\caption{Dimer binding energy versus the magnetic field. $E_b/h$ is determined as the
difference between the molecular (red squares) and the atomic (green squares) maxima in the radiofrequency transfer.
For the atomic transition we plot the deviations of the observed radio frequency transitions to the expected ones.
The data from coupled-channels model of Ref.\cite{Chiara} (blue circles) has been shifted to lower fields by 0.08~G to
match the experiment. A fit of the theoretical binding energy with the two-channel model of Ref.\cite{Pricoupenko2},
also given in Eq. \ref{Pricoupenko} (blue line), predicts $B_0=58.901(5)$~G (vertical dashed line).
The inset shows a typical molecule association spectrum from the $m_F$=-1 to the
$m_F$=0 state, where the signal is the number of atoms left in $m_F$=-1 after the radiofrequency transfer. Both atomic and
molecular signals can be seen.} \label{BE}
\end{figure}

\subsection{Measurement of the dimer binding energy and estimation of $R^*$}
We obtained another confirmation of the quality of the CC model by directly measuring the binding
energy of the weakly-bound dimers at the two Feshbach resonances. We employed a standard technique:
a weakly-interacting Bose-Einstein condensate was prepared in the $m_F$=-1 state, and then
transferred to the $m_F$=0 state by a radio-frequency pulse. At small detunings from the resonance,
we observed a transfer to both the atomic and a molecular state, as shown in the example in the
inset of Fig.\ref{BE}. The atomic signal is power broadened because the Rabi frequencies of
the atom-molecule and atom-atom transitions are very different already for small detunings. We
fitted the spectra of the atomic and molecular signals with Gaussian distributions. The molecular
binding energy $E_b$ was determined as twice the energy detuning between the atomic and molecular
peaks, taking conservatively as uncertainty the quadrature of the individual widths of the fitted
Gaussian. For example, Fig.\ref{BE} shows the measured $E_b$ for the resonance centered at
$B_0\simeq$58.9~G, for which we found a very good agreement with the CC model, within the 0.1~G
uncertainties of both the CC model and the experiment. A similar agreement was found also for the other narrow
resonance at $B_0\simeq$65.6~G.

It is known that the near-threshold molecular structure of $^{39}$K cannot be rigorously
interpreted in terms of one- or two-channel models over a wide range of magnetic field values,
because of the presence of broad avoided crossings \cite{Chiara}. This makes the calculation of the
intrinsic length through the equation $R^*= \hbar^2/ma_{bg}\delta\mu\Delta$ difficult, since the
relative molecular magnetic moment $\delta\mu$ changes with the magnetic field. The values of
$\delta\mu$ previously reported in \cite{Chiara}, represent indeed the average magnetic moment in a
finite region of magnetic fields close to the resonance centers.

To determine $R^*$ more accurately, we first extracted from our numerically calculated collision
phase shift the value $r_e^{\rm res}$ of the effective range at resonance ($a\to \infty$).  We then
used the relation between $r_e^{\rm res}$ and the intrinsic resonance length
\begin{equation}
r_e^{\rm res}=-2 R^* +\frac{2}{3 \pi} \Gamma^2 \left( \frac{1}{4} \right) R_{vdW}
\end{equation}
obtained within a quantum-defect model in Ref.\cite{Gao}.

To gain more analytical insight we also used an exactly solvable two-channel model developed in
Ref.\cite{Pricoupenko2, Pricoupenko1}, which has already been adapted to narrow and
intermediate-strength Feshbach resonances such as the ones in $^{39}$K. In this model, the binding
energy for $|a|\gg a_{bg}$ is expressed as $E_b=\hbar^2q_{dim}^2/2m$, with
\begin{equation}
q_{dim}=(a-\sqrt{a^2-2ar_e})/ar_e)\,,
\label{Pricoupenko}
\end{equation}
where the effective range is
\begin{equation}
r_e\simeq  -2R^*(1-a_{bg}/a)+4b/\sqrt{\pi}-2b^2/a \,.
\label{Pricoupenko1}
\end{equation}
Here $b$ is a fitting parameter of the order of the van der Waals radius.

We fitted the calculated binding energies for all seven Feshbach resonances investigated in the
experiment, using the calculated values for the background scattering length $a_{bg}$ and for the
resonance width $\Delta$ to evaluate $a(B)$, while leaving both $b$ and $R^*$ as fit parameters. In
the fit we employed CC data in a range of detunings $|B-B_0|<\delta B_{max}$, with a maximum
$\delta B_{max}$ of the order of $0.5\Delta$ for each resonance; we derived a characteristic
uncertainty on the fitted quantities by comparing the results for the cases $\delta
B_{max}=0.75\Delta$ and $\delta B_{max}=0.25\Delta$. As shown in Table \ref{tabsi:results}, the determined values of
$R^*$ are in very good agreement with the numerically exact CC calculation.

\begin{table*}[ht]
\centering \caption{Theoretical and experimental parameters for the Feshbach resonances in the spin
channels $m_F$: resonance centers from the CC model ($B_0^{th}$) and the experiment ($B_0^{exp}$);
resonance widths $\Delta$ and background scattering lengths $a_{bg}$ \cite{Chiara}; intrinsic
length ($R^*$) calculated from the on-resonance effective range (see text) or ($R^*_{2ch}$) derived
from a fit to a two-channel model \cite{Pricoupenko1}; resonance strengths $s_{res}$; measured
three-body parameter $a_-$ and decay parameter $\eta_-$; initial temperature $T$; initial atom
number $N$ and mean trap frequency $\bar{\omega}/2\pi$.}
\begin{tabular*}{16cm}{r r c c c c c c c c c c c}

\hline\hline $m_F$ & $B_0^{th}$(G)& $B_0^{exp}$(G) & $-\Delta$(G) & $a_{bg}$ ($a_0$) &
\hspace{3mm}$R^*$ & \hspace{3mm} $R^*_{2ch}$ & $s_{res}$ & $a_- (a_0)$ &  $\eta_-$ & $T$(nK)& N
(10$ ^{4} $) &$  \bar{\omega}/2\pi $ (Hz)\\ [2ex]

\hline 0  & 471.9 & 471.0 (4) &72&-28& 22 & 20(2)& 2.8 & 640 (100) & 0.065 (11) & 50 (5) & 5 (2) & 28\\
+1  & 402.4 & 402.6 (2) &52&-29& 22 & 22(3) & 2.8 & 690 (40)& 0.145 (12)& 90 (6)& 40 (3) & 14\\
 -1  & 33.6  & 33.64 (20) &-55&-19& 23 & 23(2) & 2.6 & 830 (140) & 0.204 (10)& 120 (7) & 80 (20)&14\\
-1 & 560.7 & 560.72 (15) &56&-29& 24 & 23(2) & 2.5 & 640 (90) & 0.22 (2)  & 20 (5)& 3 (2) & 14\\
 -1  & 162.3 & 162.35 (18) &37&-19& 59 & 59(3) & 1.1 & 730 (120) & 0.26 (5)& 40 (8)  & 16 (3) & 12\\
0  & 65.6 & 65.67 (5) &7.9&-18& 456 & 449(8) & 0.14 & 950 (250) & - & 330 (30)& 5(2)&140\\
 0  & 58.8 & 58.92 (3) &9.6&-18& 556 & 559(1) & 0.11 & 950(150) & - &  400 (80)& 7 (1)&136\\
 0\footnote{$d$-wave}  & 60.5 & 60.1 (1) &-0.6 & & & & & & & \\[1ex]
\hline \hline
\end{tabular*}
\label{tabsi:results}
\end{table*}

\subsection{Analysis of the $K_3$ measurements}
We now discuss in more detail the analysis of the loss measurements that resulted in the data of
Figs.2-3 of the main paper. We recall that to determine $K_3$ we fitted the coupled equations for
$N(t)$ and $T(t)$ at each magnetic field, obtaining the three quantities
$\Gamma=3\beta^2N_0^2K_3/\sqrt{27}T_0^3$, $N_0$ and $T_0$. The typical statistical errors are: 20\%
for $\Gamma$; 3\% for $N_0$; 10\% for $T_0$. The overall relative statistical error on $K_3$,
obtained by summing the individual contributions, typically amounts to 60\%, and is quantified by
the error bars in Figs.2-3 of the main paper.

In addition, we have two systematic sources of error on $K_3$. The first one is a 30\% uncertainty
on $N_0$, which comes from our imperfect knowledge of the actual absorption cross-section of the
atoms. The second one comes from the uncertainty in the mean trap frequency $\bar{\omega}$, which
is contained in the parameter $\beta$ ($\beta^2\propto\bar{\omega}^6$). This was measured by
exciting a sloshing motion of the atomic sample along the three trap axes, with a typical
uncertainty of 20\%. The overall systematic uncertainty is therefore as large as 200\%. This
implies that the $K_3$ scale is determined only up to a factor of about 3.

The unitarity-limited three-body rate contains instead only one experimental parameter, the
temperature: $K_3^{max}=36\sqrt{3}\pi^2\hbar^5/(k_{B}T)^2m^3$. In our analysis we used the initial
temperature $T_0$ to estimate $K_3^{max}$, with an overall uncertainty that is negligible with
respect to the scaling factor above.

When fitting the experimental data for the five broad resonances, we allowed for a single fitting
factor in front of the overall rate $(1/K_3(a)+1/K_3^{max})^{-1}$. In all cases we found factors
smaller than 3, in agreement with the estimated uncertainty.

\begin{figure}[ht]
\includegraphics[width=0.95\columnwidth] {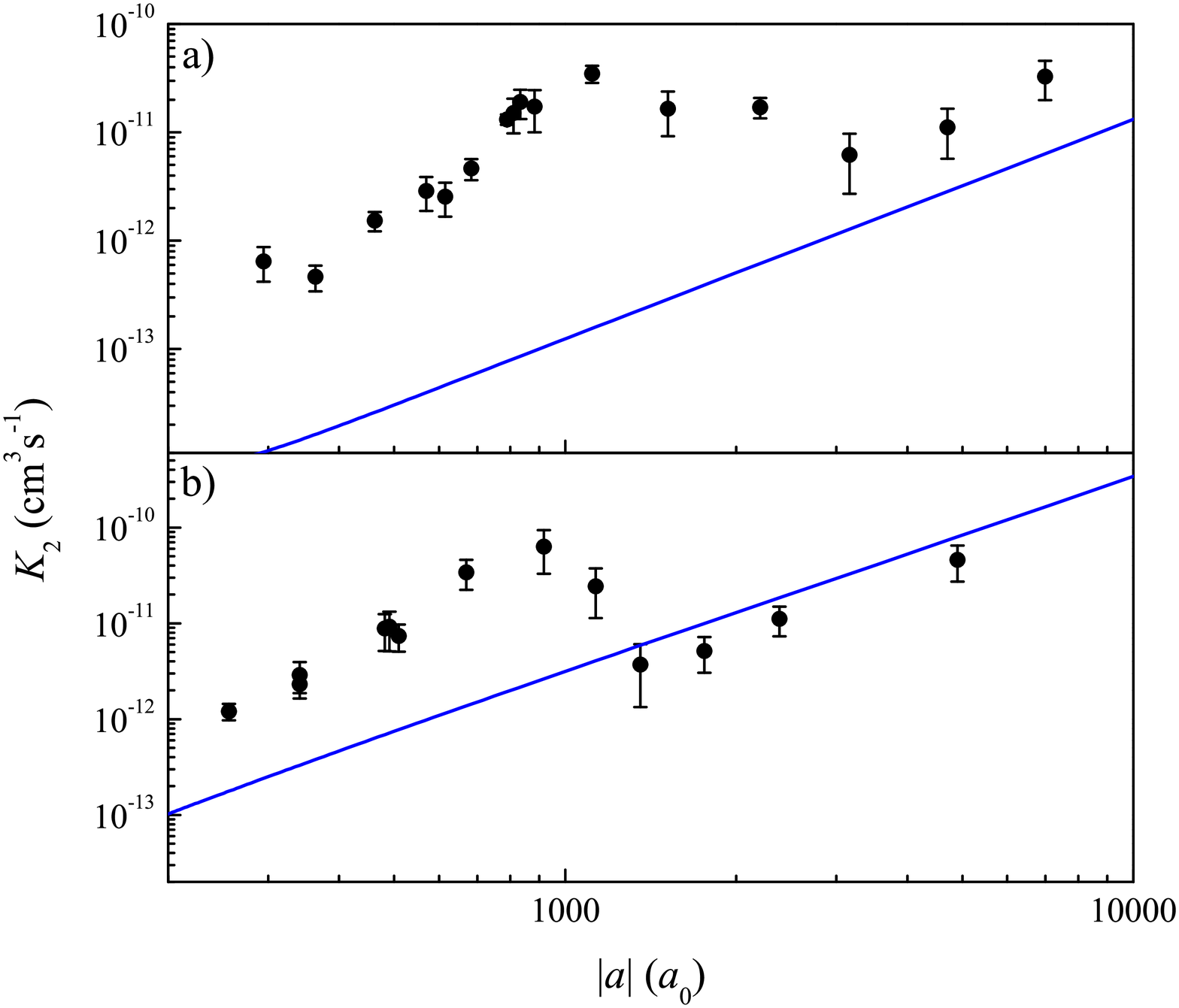}
\caption{Comparison of the measured $K_2$ (dots) with the CC calculations (solid line),
for the narrow Feshbach resonances around: a) 58.8 G; b) 65.6 G.} \label{K2ab}
\end{figure}

\subsection{Three-body and two-body decay}
In principle, atoms in the excited $m_F$=0,-1 states can decay also via two-body relaxation. The
two-body rate is normally very small, since it involves processes that change the total $m_F$, but
it can become relevant close to the Feshbach resonances, where the universal behavior for the two
body constant $K_2\propto a^2$ sets in. While the measured $K_3$ at the broader Feshbach resonances
follows the expected $a^4$ behavior, a possible reason for the slower variation of $K_3$ with $a$
at the two narrow Feshbach resonances might be indeed a contribution of two-body processes, namely
dipolar relaxation of a pair of $m_F$=0 atoms into $m_F$=1 atoms. To investigate this possibility,
we tried to model the observed evolution of $N(t)$ and $T(t)$ also as a two-body decay, for which
the rate equations are
\begin{eqnarray}
\frac{dN}{dt}&=&-K_2\langle n\rangle N\,,\\
\frac{dT}{dt}&=&\frac{K_2\langle n\rangle T}{4}\,,
\end{eqnarray}
where $\langle n\rangle=(1/N)\int n(\vec{x})^2d^3x$ is the mean density. A compact solution for the
average density can be found as
\begin{equation}
\langle n(t)\rangle=\frac{\langle n_0\rangle}{1+\frac{11}{8}\langle n_0\rangle K_2 t}\,,
\end{equation}
which is nominally different from the analogous solution for $\langle n(t)^2\rangle$ in presence of
three-body losses
\begin{equation}
\langle n(t)^2\rangle=\frac{\langle n_0^2\rangle}{1+3 \langle n_0^2\rangle K_3 t}\,.
\end{equation}
However, the two solutions are not too different, and the experimental data can in general be
fitted almost equally well using the two models.

Since $K_2$ can be predicted exactly by the CC model, we could however compare a $K_2$ fitted from
the experimental data assuming only two-body losses, with the calculated one. As shown in
Fig.\ref{K2ab}, this comparison indicates that the two-body decay is unlikely to play a role in the
observed decay for the resonance at 58.9~G, but might indeed contribute to the observed background
losses for the resonance at 65.6~G. Note however that we estimate a typical uncertainty of about 2
in the overall scale of the measured $K_2$, so that further detailed experiments, possibly with
different $T_0$ and $N_0$ will be necessary to assess the real impact of two-body losses even in
the latter case. In any case, the measured loss maxima cannot be explained at all in terms of
two-body losses. A similar analysis we performed for the five broader resonances indicates that in
those cases the two-body losses are more than one order of magnitude smaller that the three-body
ones.


\begin{thebibliography}{99}

\bibitem{Kraemer} T. Kr\"{a}mer, M. Mark, P. Waldburger, J. G. Danzl, C. Chin, B. Engeser, A. D.
    Lange, K. Pilch, A. Jaakkola, H.-C. N\"{a}gerl and R. Grimm, Nature \textbf{440}, 315
    (2006).
\bibitem{Potassium} M. Zaccanti, B. Deissler, C. D'Errico, M. Fattori, M. Jona-Lasinio, S.
    M\"{u}ller, G. Roati, M. Inguscio and G. Modugno, Nat. Phys. \textbf{5}, 586 (2009).
\bibitem{Minardi} G. Barontini, C. Weber, F. Rabatti, J. Catani, G. Thalhammer, M. Inguscio, F.
    Minardi Phys. Rev. Lett. \textbf{103}, 043201 (2009).
\bibitem{Hulet} S. E. Pollack, D. Dries, and R. G. Hulet, Science \textbf{326}, 1683 (2009).
\bibitem{Khaykovich1} N. Gross, Z. Shotan, S. Kokkelmans, and L. Khaykovich, Phys. Rev. Lett.
    \textbf{103}, 163202 (2009).
\bibitem{Khaykovich2} N. Gross, Z. Shotan, S. Kokkelmans, and L. Khaykovich, Phys. Rev. Lett.
    \textbf{105}, 103203 (2010).
\bibitem{Jochim1} T. B. Ottenstein, T. Lompe, M. Kohnen, A. N. Wenz, and S. Jochim, Phys. Rev.
    Lett. \textbf{101}, 203202 (2008).
\bibitem{Hara1} J. H. Huckans, J. R. Williams, E. L. Hazlett, R.W. Stites, and K. M. O'Hara, Phys.
    Rev. Lett. \textbf{102}, 165302 (2009).
\bibitem{Hara2} J. R. Williams, E. L. Hazlett, J. H. Huckans, R.W. Stites, Y. Zhang, and K. M.
    O'Hara, Phys. Rev. Lett. \textbf{103}, 130404 (2009).
\bibitem{Jochim2} T. Lompe, T. B. Ottenstein, F. Serwane, K. Viering, A. N. Wenz, G. ZuÂ¨rn, and S.
    Jochim, Phys. Rev. Lett. \textbf{105}, 103201 (2010).
\bibitem{Ueda} S. Nakajima, M. Horikoshi, T. Mukaiyama, P.Naidon, and M. Ueda, Phys. Rev. Lett.
    {\bf 106}, 143201 (2011).
\bibitem{Rb} R. J. Wild, P. Makotyn, J. M. Pino, E. A. Cornell, and D. S. Jin, Phys. Rev. Lett.
    \textbf{108}, 145305 (2012).
\bibitem{Hulet1} P. Dyke, S. E. Pollack, R. G. Hulet, arXiv:1302.0281v1 (2012).
\bibitem{Efimov} V. Efimov, Phys. Lett. B \textbf{33}, 563 (1970).
\bibitem{Nielsen} E. Nielsen, D. V. Fedorov, A. S. Jensen and E. Garrido, Phys. Rep. \textbf{347},
    373 (2001).
\bibitem{Braaten} E. Braaten and H.-W. Hammer, Phys. Rep. \textbf{428}, 259 (2006).
\bibitem{review} H.-W- Hammer and L. Platter, Annu. Rev. Nucl. Part. Sci. \textbf{60}, 207 (2010).
\bibitem{magnons} Y. Nishida, Y. Kato	and C. D. Batista, Nat. Phys. \textbf{9},
    93 (2013).
\bibitem{Berninger} M. Berninger, A. Zenesini, B. Huang, W. Harm, H. C. N\"{a}gerl, F. Ferlaino,
    R. Grimm, P. S. Julienne, and J. M. Hutson, Phys. Rev. Lett. \textbf{107}, 120401 (2011).
\bibitem{Chin} C. Chin, arXiv:1111.1484v2 (2011).
\bibitem{Greene} J. Wang, J. P. D'Incao, B. D. Esry, and Chris H. Greene, Phys. Rev. Lett.
    \textbf{108}, 263001 (2012).
\bibitem{Ueda1} P. Naidon, S. Endo, and M. Ueda, arXiv:1208.3912v1 (2012).
\bibitem{Feshbach} C. Chin, R. Grimm, P. S. Julienne, and E. Tiesinga, Rev. Mod. Phys. \textbf{82},
    1225 (2010).
\bibitem{Petrov} D. S. Petrov, Phys. Rev. Lett. \textbf{93}, 143201 (2004).
\bibitem{Gogolin} A. O. Gogolin, C. Mora and R. Egger, Phys. Rev. Lett. \textbf{100}, 140404
    (2008).
\bibitem{Massignan} P. Massignan and H. T. C. Stoof, Phys. Rev. A, \textbf{78} 030701(R) (2008) .
\bibitem{Pricoupenko1} M. Jona-Lasinio and L. Pricoupenko, Phys. Rev. Lett. \textbf{104}, 023201
    (2010).
\bibitem{Esry} Y. Wang, J. P. D'Incao and B. D. Esry, Phys. Rev. A \textbf{83}, 042710 (2011).
\bibitem{Pricoupenko2} L. Pricoupenko and M. Jona-Lasinio, Phys. Rev. A, \textbf{84}, 062712
    (2011).
\bibitem{Zwerger} R. Schmidt, S. P. Rath and W. Zwerger, Eur. Phys. J. B \textbf{85}, 386 (2012).
\bibitem{Zinner} P. K. Sorensen, D. V. Fedorov, A. S. Jensen, N. T. Zinner, Phys. Rev. A {\bf 86},
    052516 (2012).
\bibitem{Chiara} C. D'Errico, M. Zaccanti, M. Fattori, G. Roati, M. Inguscio, G. Modugno and A.
    Simoni, New J. Phys. {\bf 9} , 223 (2007).
\bibitem{BEC39} M. Landini, S. Roy, G. Roati, A. Simoni, M. Inguscio, G. Modugno and M. Fattori,
    Phys. Rev. A {\bf 86}, 033421 (2012).
\bibitem{Weber} T. Weber, J. Herbig, M. Mark, H-C. N\"{a}gerl, and R. Grimm, Phys. Rev. Lett.
    \textbf{91}, 123201 (2003).
\bibitem{suppl} See the supplementary material for more details on the measurements and their
    modelling.
\bibitem{Grimm} F. Ferlaino, A. Zenesini, M. Berninger, B. Huang, H.-C. N\"{a}gerl, R. Grimm,
    Few-Body Syst. \textbf{51}, 113 (2011).
\bibitem{unitary} J. P. D'Incao, H. Suno, and B. D. Esry, Phys. Rev. Lett. \textbf{93}, 123201
    (2004).
\bibitem{Petrov1} B. S. Rem, A. T. Grier, I. Ferrier-Barbut, U. Eismann, T. Langen, N. Navon, L.
    Khaykovich, F. Werner, D. S. Petrov, F. Chevy, and C. Salomon, arXiv:1212.5274v2 (2012).
\bibitem{Gao} B. Gao, Phys. Rev. A \textbf{84}, 022706 (2011).

\end{thebibliography}
\end{document}